1

# Nano-sculptured thin film thickness variation with  incidence angle

## Cristina Buzea* and Kevin Robbie

Department of Physics & Centre for Manufacturing of Advanced Ceramics and Nanomaterials, Queen's University, Kingston, Ontario, Canada K7L 3N6

### Abstract

*In-situ monitoring and calibration of nano-sculptured thin film thickness is a critical problem due to substrate tilt angle dependent porosity and mass flux. In this letter we present an analytical model for thickness dependence on fabrication parameters for nano-sculptured films. The generality of the model includes universal Gaussian-type flux distribution, non-unity sticking coefficients, variable off-axis sensor location, and substrate tilt. The resulting equation fits well the experimental data. The results can be particularized for films deposited at normal incidence.*

### 1. Introduction

Nano-sculptured thin films are a new class of films deposited on substrates with controlled azimuthal rotation, $\omega$, and tilt, $\theta$, by a method called glancing angle deposition (GLAD) [1], [2], [3]. The understanding and modeling [4], [5], [6] of nano-sculptured films deposited on tilted substrates become increasingly important as their applications encompass various disciplines: photonics [7], [8], [9], [10], liquid crystal display technology [11], magnetic media information storage [12], organic or inorganic sensors [13], energy storage technology [1], among others. Examples of nanostructures obtainable with GLAD are shown in Fig. 1. Thickness calibration is a common problem encountered in the case of these films primarily as a result of substrate tilt angle dependent porosity and flux capture.

In this paper we give a qualitative and quantitative description of thin film thickness calibration dependence on deposition parameters for nano-sculptured thin films fabricated at glancing angle incidence. The experimental data for nano-pillar thin films are well fitted by our equation. The results of this study can be particularized for films fabricated at normal angle incidence.

Two competing mechanisms occur for increasing substrate tilt angle for a given incident mass flux density. On one hand, film porosity increases, leading to larger film thickness. On the other hand, flux capture decreases, leading to diminished film thickness. The initial stage of material nucleation and growth on the substrate is critical for the resulting morphology. Geometric shadowing (due to substrate tilt) and low adatom mobility causes the appearance of nanostructures with large porosity. High levels of porosity affect the film thickness, according to the mass conservation law. If the substrate tilt θ is kept fixed during deposition, the structure of the resulting film consists of nanocolumns inclined towards the evaporation source (Fig. 2b). In the case of films fabricated on substrates tilted at variable tilt angles, the resulting morphology consists of bent columns which follow the direction of incoming vapors (Fig. 2c). If a rapid azimuthal rotation accompanies the substrate tilt, the resulting nanostructure is composed of an array of pillars normal to the substrate (Fig. 2d). In addition to the above typical thin film morphologies, various other can be fabricated with the GLAD process [1]. Thin film porosity depends strongly on substrate tilt angle. In the case of thin films fabricated with incoming vapor direction parallel to substrate normal (Fig. 2a), the porosity will be minimal, with density comparable to bulk values.

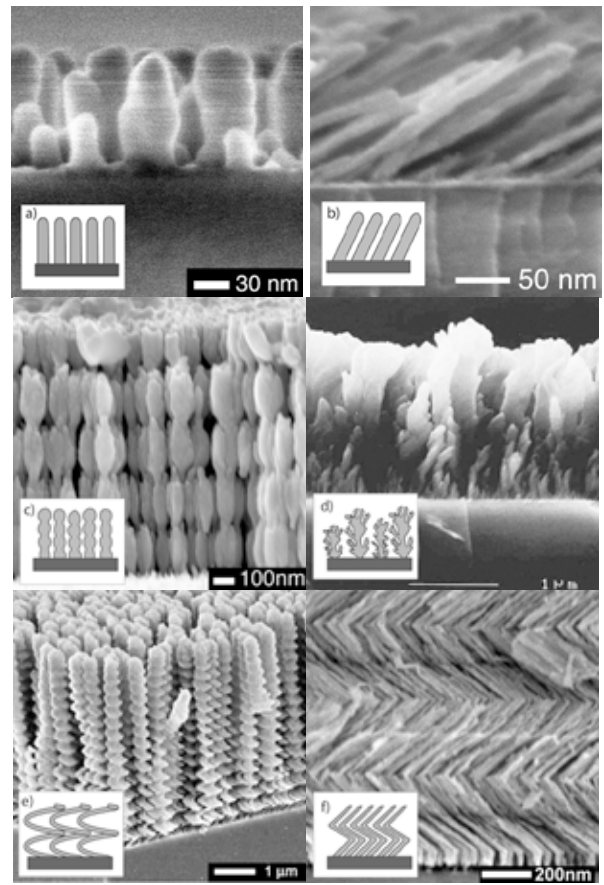

**Figure 1.** Cross section of scanning electron images showing examples of nanostructures fabricated with GLAD a, b [10], c [1], d [6], e [3], [11], f [1].

* e-mail cristi@physics.queensu.ca

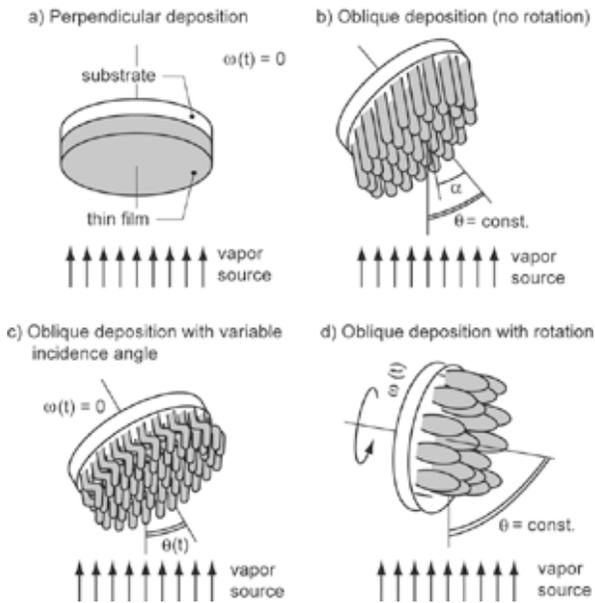

**Figure 2.** Schematics of GLAD geometry. a) Deposition with normal incidence of vapor flux, resulting in non-porous thin film. b), c), d) Oblique deposition at a fixed or variable angle without or with azimuthal substrate rotation. The columns tilt angle α is always smaller than substrate tilt angle θ.

Moreover, at oblique incidence, when the substrate normal makes an angle with the deposition direction, the amount of incoming particles which are captured by the substrate will be smaller than in the case of normal incidence, due to geometrical considerations.

Usually, the sensors used in monitoring the thickness and deposition rate of nano-sculptured thin films introduce large errors as they are unable to account for the effects of the tilt angle and porosity. Consequently, thin film deposition rate and thickness measured via various sensors transferred directly to oblique films with substrate motion result in significant error in estimating growth rates and film thickness. For example, the error in thickness reading of silicon thin films deposited on tilted substrate with constant azimuthal substrate rotation can be more than 50% of the nominal thickness [14].

This study has significant implications for in-situ monitoring of deposition rate and thickness for thin films fabricated on tilted substrates by evaporation techniques, sputtering, or laser ablation.

Ideally an in-situ deposition rate monitoring technique should determine the actual film growth rate on the substrate, or at least in its vicinity. For the sensors which measure non-porous film thickness directly on the substrate there is no need for calibration (as in the case of spectroscopic ellipsometry or multiple wavelength pyrometric interferometry). In the case of in-situ flux measurement between the deposition source and substrate surface (atomic absorption spectroscopy), a calibration related to the angle of incidence is necessary. In the general case and most common, when the sensor is situated off-axis compared to the substrate-source axis (quartz crystal microbalance or optical fibre sensor), calibration pertinent to the geometrical distribution of the vapors is necessary. More exactly, the sensor will measure the thickness of a film with low porosity at normal vapor incidence on the sensor (as the one depicted in Fig. 2a), while the deposited film will have high porosity and a different morphology (Fig. 2b, c or d).

## 2. Model

In the following we calculate the thickness dependence on deposition parameters for thin films deposited at oblique incidence of vapors compared to substrate normal. The areal flux of particles of mass

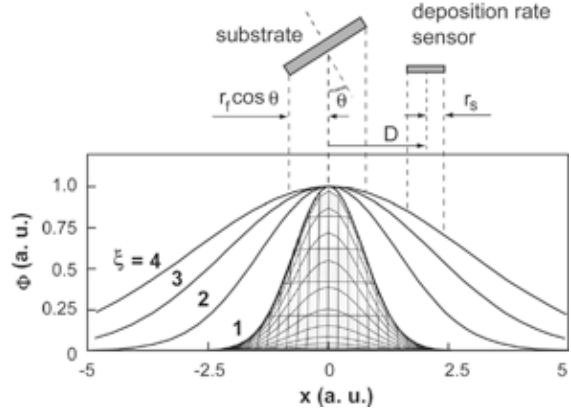

**Figure 3.** The normalized Gaussian flux distribution $\Phi_G$ variation for different values of the tapering coefficient, $\xi$. The upper portion shows the coordinates of the substrate and the off-axis deposition rate sensor relative to the Gaussian vapor distribution.

dM passing the normal unit area, $dS_n$, in the unit time, dt, is

$$\Phi = \frac{dM}{dt \cdot dS_n}. \qquad (1)$$

In the case of sensor probing position situated between the source of deposition and substrate, the amount of flux on the film is equal to the amount of flux on the sensor. Let's assume the films are deposited onto tilted substrates, their normal making an angle $\theta$ with the direction of incoming vapors. From equating the flux on the sensor, $\rho_b \cdot \delta_r / S_c$, to the flux on the film, $\rho_f H_f / (dt \cdot S_{cf} \cdot \cos\theta)$, one obtain the dependence of film thickness on film density and deposition conditions

$$\rho_f \cdot H_f = C \cdot S_{cf} \cdot \cos\theta, \qquad (2)$$

where C is a constant depending on the deposition parameters

$$C = \frac{\rho_b \cdot \delta_r \cdot dt}{S_c}. \qquad (3)$$

In Eqs. (2) and (3) we used the notations $\rho_f$ - film density, $H_f$ - film thickness, $S_{cf}$, $S_c$ - sticking coefficient on the film and sensor, respectively, dt - duration of deposition, $\rho_b$ - density of the bulk material, $\delta_r$ - the as measured deposition rate defined as the thickness of material onto the sensor divided by the duration of deposition. The sticking coefficient is defined as the number of captured atoms/molecules divided by the total number of incident atoms/molecules.

In the case of laterally located sensors compared to the substrate position, due to the non-negligible dimensions of film and sensor as well as film-sensor separation, one must take into account the geometry of the evaporated particle flux spatial distribution. In the general case a satisfactory description of spatial distribution generated by an evaporating crucible is a Gaussian-type flux, $\Phi_G$,

$$\Phi_G = \Phi_0 \cdot \exp\left(-\frac{x^2}{\xi^2}\right), \qquad (4)$$

where $\Phi_0$ is the flux peak amplitude, and $\xi$ is the Gaussian tapering coefficient, both functions of material and deposition rate. More precisely, $\xi$ describes the width of the Gaussian function, as shown in Fig. 3. It may be derived as a function of boat design, collimator aperture, deposition geometry, and process conditions. Evaporated atoms with high kinetic energy are more likely to have lower $\xi$ values with narrower distribution of flux, while slower species will be characterized by higher $\xi$ values with wider Gaussian distributions. As a result of the Gaussian distribution, the number of particles arriving onto the deposition rate sensor situated off-axis will be less than onto the substrate, which is located in the region of maximum flux, schematically illustrated in Figure 3.

We equate Eqs. (1) and (4), express the area as $dS_n = L \cdot dx$, and integrate between spatial coordinates $\pm r_f \cdot \cos\theta$, obtaining the flux arriving on the film surface, $\Phi_f$

$$\frac{H_f \cdot \rho_f}{dt \cdot S_{cf}} = \Phi_0 \cdot \frac{\xi \cdot \sqrt{\pi}}{2 \cdot r_f} \cdot Erf\left[\frac{r_f \cdot \cos\theta}{\xi}\right], \quad (5)$$

where $r_f$ - is the film radius, $\theta$ - the substrate tilt (Fig. 3), and *Erf* is the error function.

The flux of particles measured by an off-axis sensor, $\Phi_S$, can be derived in a similar manner as being

$$\frac{\delta_r \cdot \rho_b}{S_c} = \Phi_0 \cdot \frac{\xi \cdot \sqrt{\pi}}{4 \cdot r_s} \cdot \left( Erf\left[\frac{D+r_s}{\xi}\right] - Erf\left[\frac{D-r_s}{\xi}\right] \right) \quad (6)$$

where in the integration limits we have taken into account the fact that the sensor of radius $r_s$ is situated at a distance $D$ relative to the symmetry axis of the Gaussian flux.

In order to relate the film density and thickness to deposition parameters - tilt angle, sticking coefficients, sensor and substrate position with respect to the Gaussian distribution of vapor flux, we insert $\Phi_0$ from Eq. (6) into (5) obtaining

$$\rho_f \cdot H_f = 2 \cdot C \cdot S_{cf} \cdot \frac{\dfrac{r_s}{r_f} \cdot Erf\left[\dfrac{r_f \cdot \cos\theta}{\xi}\right]}{Erf\left[\dfrac{D+r_s}{\xi}\right] - Erf\left[\dfrac{D-r_s}{\xi}\right]} \quad (7)$$

The significance of Eq. (7) resides in the connection between film thickness and porosity, the two being inseparable quantities affected by the substrate tilt angle $\theta$. In accordance to Eq. (7), the thickness of the deposited film depends strongly on the film porosity, deposition geometry and deposition parameters. The information of the substrate tilt angle is contain in the Error function argument.

Recent simulation studies concluded that the sticking coefficient depends on the angle on arrival onto the substrate [15], in our case on the substrate tilt $\theta$. In order to describe qualitatively the behavior of the sticking coefficient with vapor incidence angle $\theta$ and particle incident energy, we propose an empirical equation which fits the simulation results [15]

$$S_{cf}(\theta) = 1 - \exp\left(-\frac{(\theta - \alpha)^2 + y^2}{\mu^2}\right). \quad (8)$$

## 3. Comparison with experimental data

In Fig. 4 we compare the deduced equations for thickness and porosity with experimental data for silicon thin films deposited on tilted substrates with rapid azimuthal rotation, and a laterally located deposition rate sensor [14]. One notices in Fig. 4 that Eq. (7) alone can roughly fit the experimental data.

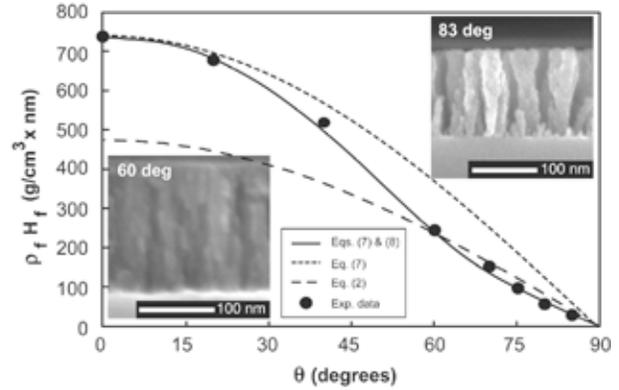

**Figure 4.** Comparison of experimental data [14] of the product film density thickness, $\rho_f \cdot H_f$, with theoretical curves in the case of constant areal flux given by Eq. (2), Gaussian flux - Eq. (7), and Gaussian flux together with angle dependent sticking coefficient Eqs. (7) & (8) with the fitting parameters $\delta = 6.23$, $\alpha = 70$, $\mu = 25$, and y = 30. The deposition parameters are $\rho_b = 2.33$ g/cm$^3$, $\delta_r = 0.5$ nm/s, dt = 408 s, $r_f = 1.27$ cm, $r_s = 0.7$ cm, and D = 4.2 cm. Inset are scanning electron microscopy images of lateral view of films fabricated at 60 and 83 incidence angle [10].

However, the best fit is given by Eqs. (7) and (8) taking into account an angle dependent sticking coefficient, with fitting parameters given in the caption of Fig. 4. On the other side, Eq. (2) cannot describe the experimental data for films deposited at either oblique or normal incidence ($\theta$=0), being valid only for sensors probing location situated on-axis, between the substrate and source.

One can conclude from Eq. (7) that an accurate *in-situ* thickness calibration of thin films fabricated on tilted substrates is possible only by knowing their density value which varies with the incidence angle, as shown by experimental data [16], [17] and simulations [18]. It has been suggested that the density variation with tilt is essentially the same for various morphologies of the columns manufactured of the same material: zig-zag, helicoidal or cylindrical [19], however, to date no consensus exists upon a quantitative description (general equation) of density variation with substrate tilt. Therefore, *in-situ* thickness calibration of nano-sculptured thin films requires a prior experimental determination of the film density variation with the incidence angle.

## 4. Conclusions

In conclusion, we calculated the relation between nano-sculptured thin film thickness and porosity on substrate tilt assuming a Gaussian-type amplitude of the evaporated particles flux. The generality of the equation is given by its dependence on non-unity sticking coefficients, substrate and thickness sensor physical dimensions and positions, and deposition conditions (rate and duration). Our results are in very good agreement with experimental data, and can also be particularized for films deposited at normal incidence.

## Acknowledgments

We gratefully acknowledge financial support from the Natural Sciences and Engineering Research Council of Canada (NSERC), the Canada Research Chairs Program (CRC), and the Canadian Institute for Photonic Innovations (CIPI).